\documentclass[useAMS,usenatbib]{mn2e}

\usepackage{graphicx}
\usepackage{txfonts}
\usepackage{natbib}
\newcommand\aj{AJ}
\newcommand\apj{ApJ}

\newcommand\aap{A\&A}
\newcommand\mnras{MNRAS}
\newcommand\apjl{ApJ}
\newcommand\pasp{PASP}
\newcommand\nat{Nature}

\title[]{The M\,4 Core Project with {\it HST} -- II. Multiple Stellar Populations at the Bottom of the Main Sequence.}
\author[A.\, P.\, Milone et al.]
{A.\,P.\,Milone$^{1}$, A.\,F.\,Marino$^{1}$, L.\,R.\,Bedin$^{2}$, G.\,Piotto$^{3,2}$,  S.\,Cassisi$^{4}$, A.\,Dieball$^{5}$, 
\newauthor J.\,Anderson$^{6}$, H.\,Jerjen$^{1}$, M.\,Asplund$^{1}$, A.\,Bellini$^{6}$, K.\,Brogaard$^{7,8}$, A.\,Dotter$^1$,  
\newauthor M.\,Giersz$^{9}$, D.\,C.\,Heggie$^{10}$, C.\,Knigge$^{5}$, R.\,M.\,Rich$^{11}$, M.\,van den Berg$^{12,13}$, R.\,Buonanno$^{4}$ \\
$^{1}$Research School of Astronomy \& Astrophysics, Australian National University, Mt Stromlo Observatory, via Cotter Rd, Weston, ACT 2611, Australia \\
$^{2}$Istituto Nazionale di Astrofisica - Osservatorio Astronomico di Padova, Vicolo dell'Osservatorio 5, Padova, IT-35122\\
$^{3}$Dipartimento di Fisica e Astronomia ``Galileo Galilei'', Univ. di Padova, Vicolo dell'Osservatorio 3, Padova IT-35122\\
$^{4}$INAF-Osservatorio Astronomico di Collurania, via Mentore Maggini, I-64100 Teramo, Italy\\
$^{5}$Department of Physics and Astronomy, University Southampton, Southampton SO17 1BJ, UK\\
$^{6}$Space Telescope Science Institute, 3800 San Martin Drive, Baltimore,  MD 21218, USA\\
$^{7}$Department of Physics and Astronomy, Aarhus University, Ny Munkegade, 8000 Aarhus C, Denmark\\
$^{8}$Department of Physics and Astronomy, University of Victoria, PO Box 3055, Victoria, B.C.,V8W 3P6, Canada\\
$^{9}$Nicolaus Copernicus Astronomical Center, Polish Academy of Sciences, ul. Bartycka 18, 00-716, Warsaw, Poland\\
$^{10}$School of Mathematics and Maxwell Institute for Mathematical Sciences, University of Edinburgh, Kings Buildings, Edinburgh, UK-EH9-3JZ\\
$^{11}$Department of Physics and Astronomy, University of California, Los Angeles, CA 90095, USA\\
$^{12}$Astronomical Institute ``Anton Pannekoek'', University of Amsterdam, Science Park 904, 1098 XH Amsterdam, The Netherlands\\
$^{13}$Harvard-Smithsonian Center for Astrophysics, 60 Garden Street, Cambridge, 02138 MA, USA\\
 }

\begin{document}

\pagerange{\pageref{firstpage}--\pageref{lastpage}} \pubyear{2013}

\maketitle
\label{firstpage}

\begin{abstract}  
The M\,4 Core Project with {\it HST} is designed to exploit the {\it Hubble Space Telescope} to investigate the central regions of M\,4, 
the Globular Cluster closest to the Sun. 
In this paper we combine optical and near-infrared photometry to study multiple stellar populations in M\,4.  
We detected two sequences of M-dwarfs containing $\sim$38\% ($MS_{\rm I}$) and $\sim$62\% ($MS_{\rm II}$) of MS stars below the main-sequence (MS) knee. 
We compare our observations with those of NGC\,2808, which is the only other GCs where  multiple MSs of very low-mass stars have been studied to date.
  We calculate synthetic spectra for M-dwarfs, assuming the chemical composition mixture inferred from spectroscopic studies of stellar populations along the red giant branch, and different Helium abundances,
        and we compare predicted and observed colors. 
        Observations
        are consistent with two populations, one with primordial abundance and another with enhanced nitrogen and depleted oxygen. 
\end{abstract}


\section{Introduction}\label{sec:intro}
 The M\,4 Core Project with {\it HST} is a {\it Hubble Space Telescope} ({\it HST}) large program (GO-12911, PI: Bedin) designed to provide high-accuracy astrometry and photometry for stars in the core of the closest Globular Cluster (GC) M\,4 ($d_\odot$=2.2\,kpc, Harris\,1996 updated as in 2010).
The observations are designed to detect binary 
dark companions (such as a black hole, a white dwarf, or a neutron star)  of bright MS stars
by measuring the `wobble' of the luminous star around the center of mass of the binary system.    

 The exquisite photometry and astrometry provided by {\it HST} makes this dataset ideal for several other astrophysical applications including the search for variable stars or transiting exoplanets, the measure of absolute proper motions and annual parallax, the study of internal dynamics and of photometric binaries, and the search for intermediate-mass central black hole (see Bedin et al.\ 2013, hereafter paper I, for an overview of the project).
One of the main goals of the project is the investigation of multiple stellar populations in this cluster.

Several papers have been dedicated to the study of the different stellar generations host by
M\,4. 
Spectroscopy has revealed that M\,4 stars have inhomogeneous content of C, N, O, Al, and Na  (e.g.\, Gratton et al.\ 1986, Wallerstein 1992, Drake et al.\, 1992, Smith et al.\, 2005) and exhibit Na-O, Al-O, and C-N anticorrelations (e.g.\, Ivans et al.\, 1999, Marino et al.\, 2008, Carretta et al.\, 2009, Villanova \& Geisler 2011).  
 The distribution of [Na/Fe] and [O/Fe] is bimodal, and the two groups of Na-poor/O-rich and Na-rich/O-poor stars, which are commonly associated with a first and a second stellar generation, populate two 
distinct red-giant branches (RGBs, Marino et al.\,2008). 
 The color distribution of horizontal-branch (HB) stars is also bimodal and is strictly related to the stellar populations: blue-HB stars share the same chemical composition as second-generation stars, while the red-HB stars are consistent with being first generation (Marino et al.\, 2011).
 Photometric evidence 
of multiple populations in M\,4 (Marino et al.\,2008, Lee et al.\,2009, Monelli et al.\,2013) is based on the analysis of the RGB, while, so far, no evidence of multiple or extended MS has been found.

In a recent study of the massive GC NGC\,2808,
Milone et al.\, (2012a, hereafter M12a) have demonstrated that the F110W and F160W filters of the Infrared Channel of the Wide Field Camera 3 (WFC3/IR) on the {\it Hubble Space Telescope} ({\it HST}) are powerful tools to identify multiple sequences along the MS of very-low mass (VLM) stars.
The near-infrared (NIR) portion of the spectrum of cold M-dwarfs is very sensitive to the effects of molecules, in particular $\rm H_{2}O$. 
For this reason, NIR photometry can provide unique information on multiple populations among VLM stars.
In this paper we extend the study by M12a on NGC\,2808 to the less-massive GC M\,4. We exploit the dataset available from GO-12911 and from the {\it HST} archive to detect, for the first time, multiple stellar populations at the bottom of the MS of this GC.

\section{Data and data analysis}
 The dataset used in this paper consists in images obtained with the IR and Ultraviolet and
visual camera (UVIS) of WFC3 and the Wide Field Channel of the Advanced Camera for Surveys (WFC/ACS) of {\it HST}. In addition, we used the photometric catalogs in F606W and F775W compiled
by Bedin et al.\ (2009) in their study of the M\,4
white-dwarf cooling sequence (GO-10146, PI.\,L.\,Bedin), and the photometric catalog in F467M described in paper\,I.

WFC3/IR data are archival {\it HST} images from program GO-12602 (PI.\,A.\,Dieball). 
This data set consists of 16$\times$652s exposures through F110W and 8$\times$652s through F160W of
a $\sim$2$\times$2 arcmin field located at about 1.5 arcmin North-East from the cluster center.  
  Since the original goals of GO-12602 project were
the search for brown dwarfs in GCs  and the study of the hydrogen-burning limit and the stellar/sub-stellar border of the MS, this dataset is optimized to get high-accuracy photometry of faint M-dwarfs. 
We also used WFC3/UVIS 3$\times$911s$+$3$\times$922s exposures in F390W from GO-12602 (PI.\,A.\,Dieball) and 4$\times$50s exposures in F814W from GO-12311 (PI.\,G.\,Piotto). 
We corrected the poor charge-transfer efficiency (CTE) in the WFC3/UVIS
data by using the procedure by Anderson \& Bedin\,(2010).

Photometry and astrometry of WFC/IR images have been carried out by using a software package that is based largely on the tools presented in Anderson \& King (2006). We corrected stellar positions for geometric distortion by using the solution developed by one of us (J.\,A.). Photometry has been calibrated as in Bedin et al.\,(2005) by using online estimates for zero points and encircled energies\footnote{http://www.stsci.edu/hst/wfc3/phot\_zp\_lbn}. Both the software package and geometric distortion correction tool will be presented in a forthcoming paper. Photometry and astrometry of WFC3/UVIS have been determined as in paper\,I.

The CMDs have been corrected for differential reddening as in Milone et al.\ (2012b). 
First, we selected a subsample of stars by including only those cluster members that lie on the MS and are brighter than the the knee. We extracted the MS ridge line and for each star, we estimated
how the 35 neighbouring stars in the subsample systematically lie to
the red or the blue of the fiducial sequence. This systematic color
and magnitude offset, measured along the reddening line, is indicative
of the differential reddening.
 We found that the reddening variation, within the field of view analyzed in this paper is smaller than $\Delta E(B-V)=0.025$ mag,  and is consistent with the degree of differential reddening measured by Hendricks et al.\,(2012) and Monelli et al.\,(2013) in the same region.

Stellar  proper motions are determined by comparing the positions of stars
measured at two different epochs from GO-10146 ACS/WFC data (epoch: 2004.6) 
 and GO-12602 IR/WFC3 data (epoch: 2012.3), and by  following a method that has been comprehensively described in other papers (e.g.\,Bedin et al.\,2003, 2008, Anderson \& van der Marel\,2010).

\section{Multiple Sequences of M-dwarfs}\label{sec:mpop}
\begin{centering}
\begin{figure}
 \includegraphics[width=8.5cm]{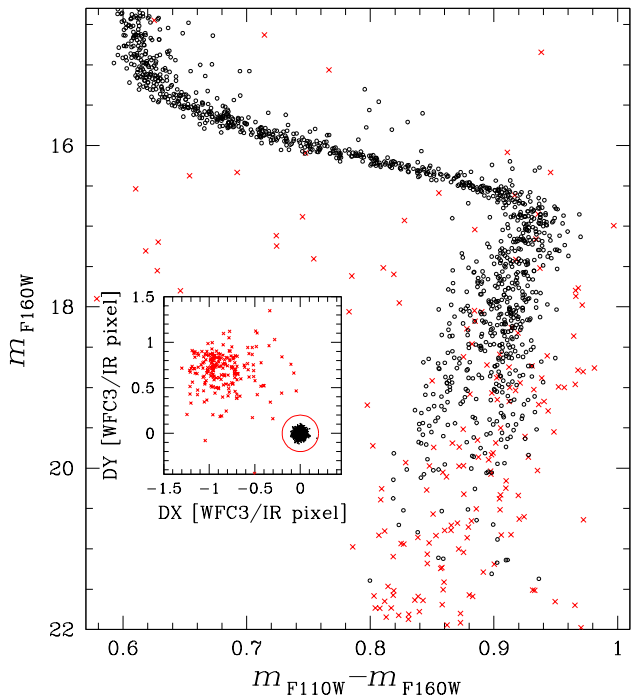}
 \caption{$m_{\rm F160W}$ versus $m_{\rm F110W}-m_{\rm F160W}$ CMD of stars in the WFC3/IR field of view. Stars that are M\,4 members according to their proper motion are plotted as black points, while field stars are represented with red crosses. The inset shows the vector-point diagram (in WFC3/IR pixel units) 
of the proper motions for the same stars as shown in the CMD. The red circle is adopted to separate cluster members from field stars.}
 \label{fig:cmd}
\end{figure}
\end{centering}

The $m_{\rm F160W}$ versus $m_{\rm F110W}-m_{\rm F160W}$ CMD (hereafter NIR CMD), corrected for differential reddening, is shown in Fig.~\ref{fig:cmd} for all the stars in the field of view. 
 The inset shows the vector-point diagram (VPD) of proper motions in WFC3/IR pixel units. 
The fact that the absolute proper motion of M\,4 differs significantly from that of background/foreground stars makes proper motions measurements the tool of choice for separating field stars from cluster members (see Bedin et al.\, 2003).

In Fig.~\ref{fig:cmd} we represent with black circles the stars that, according to their proper motion, are probable cluster members, while red crosses indicate field stars.
A visual inspection of the cluster CMD reveals that below the turn off, over a range of $\sim$2.5 F160W magnitudes ($\sim14.0 <m_{\rm F160W}< \sim17.5$), the MS is narrow and there are no evident signs of any large color spread or split. At fainter luminosities, below the MS knee, there is a dramatic change in the color distribution of MS stars, and the MS splits into two sequences. 
This section is dedicated to the analysis of these sequences of VLM stars, and we start by demonstrating that the bimodal color distribution in not entirely due to photometric errors.

To do this, an effective test consists of subdividing the available F110W, F160W, F606W and F775W images
of the same field into two independent groups (NIR and visual) and compare their CMDs.
If the color broadening is entirely due to photometric errors, a star that has bluer (or redder) colors in one CMD will have an equal probability of being red or blue in the second one. 
However, if the two sequences identified in the first data set have systematically different 
magnitudes and colors in the second one, the color broadening of the MS is more likely to be intrinsic. 
This procedure was introduced by Anderson et al.\,(2009) in their analysis of the color spread among MS stars in 47\,Tuc and is applied in the following to the case of M\,4.

\begin{centering}
\begin{figure*}
 \includegraphics[width=11cm]{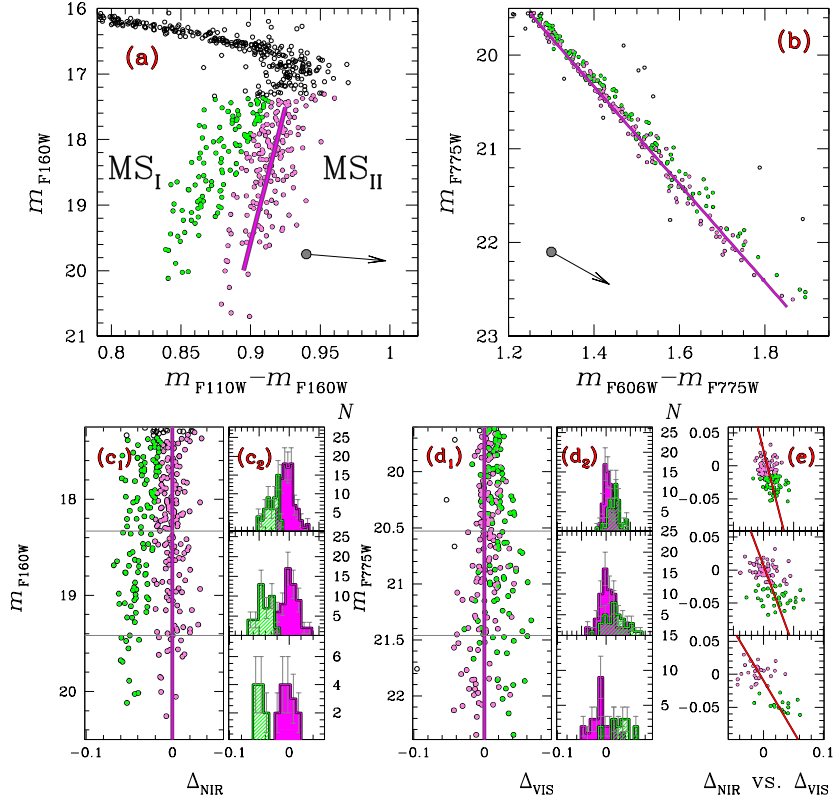}
 \caption{\textit {Upper panels}: NIR (a) and visual (b) CMD of faint MS stars with the $MS_{\rm II}$ fiducial lines colored magenta. Each star is shown in the color assigned in the NIR CMD. Arrows indicate the reddening vector and correspond in size to a reddening of E(B$-$V)=0.17. 
\textit{Panels ($\rm c$)}: Verticalized NIR CMDs ($\rm c_{1}$) and histogram distributions of $\Delta_{\rm NIR}$($\rm c_{2}$)
color residuals for $MS_{\rm I}$ (green) and $MS_{\rm II}$ (magenta) in the three $m_{\rm F160W}$ intervals  delimited by horizontal lines. 
 The verticalized visual CMDs and the histogram distributions of $\Delta_{\rm VIS}$ for MS stars in the three $m_{\rm F775W}$ intervals  delimited by horizontal lines are plotted in panels ($\rm d_{1}$) and ($\rm d_{2}$), respectively. Panel (e) shows $\Delta_{\rm NIR}$ against $\Delta_{\rm VIS}$ for stars in three F775W luminosity intervals. Least-squares best-fitting straight lines are colored red.
 }
 \label{fig:visual}
\end{figure*}
\end{centering}

In Fig.~\ref{fig:visual}, we compare two CMDs of the
faint MS stars obtained from two independent datasets. 
The NIR CMD is plotted in the upper-left panel and the $m_{\rm F775W}$ versus $m_{\rm F606W}-m_{\rm F775W}$ CMD (hereafter visual CMD) is shown in the upper-right panel. Reddening direction is indicated in each CMD by gray arrows. The length of each arrow corresponds to a reddening variation of $\Delta E(B-V)$=0.17 mag that is about seven time larger than the maximum reddening variation observed in the M\,4 field studied in this paper.

In the NIR CMD we have defined, by eye, two groups of $MS_{\rm I}$ 
and $MS_{\rm II}$  stars with $m_{\rm F160W}>17.2$ that we have colored green and magenta, respectively. 
In the visual CMD, we kept for each star the same color as defined in NIR: it is quite clear that the two groups of stars remain well separated.  Moreover, the evidence that $MS_{\rm I}$ stars have, on average redder  $m_{\rm F606W}-m_{\rm F775W}$ color than $MS_{\rm II}$ stars but bluer $m_{\rm F110W}-m_{\rm F160W}$ color
demonstrates that the color spread cannot be due to residual differential reddening.

In order to strengthen this result, in each CMD we have derived
a fiducial line of $MS_{\rm II}$ stars, which is represented as a continuous magenta line in Fig.~\ref{fig:visual}. We than calculated the differences between the $m_{\rm F606W}-m_{\rm F775W}$ (and $m_{\rm F110W}-m_{\rm F160W}$) colors of each star and the color of the fiducial at the same $m_{\rm F775W}$ (or $m_{\rm F160W}$) magnitude. These color residuals are designated $\Delta_{\rm VIS}=\Delta$($m_{\rm F606W}-m_{\rm F775W}$) and  $\Delta_{\rm NIR}=\Delta$($m_{\rm F110W}-m_{\rm F160W}$).

The verticalized $m_{\rm F160W}$ versus $\Delta_{\rm NIR}$ and  $m_{\rm F775W}$ versus $\Delta_{\rm VIS}$ CMDs are plotted in the panels ($\rm c_{1}$) and ($\rm d_{1}$), respectively.  Panels ($\rm c_{2}$) and ($\rm d_{2}$) show the corresponding histogram of the color-residual distribution for the straightened $MS_{\rm I}$ and $MS_{\rm II}$ in three luminosity intervals. In addition, in panel (e) we show  the anti-correlation between $\Delta_{\rm VIS}$ and $\Delta_{\rm NIR}$ color residuals. 
$MS_{\rm II}$ stars have on average bluer $m_{\rm F606W}-m_{\rm F775W}$ colors than $MS_{\rm I}$ stars with the same F775W magnitude. The color order is reversed in the NIR CMD. 
This behaviour further confirms that the color broadening is an intrinsic feature of the faint MS of M\,4.

%

\begin{centering}
\begin{figure*}
 \includegraphics[width=9.5cm]{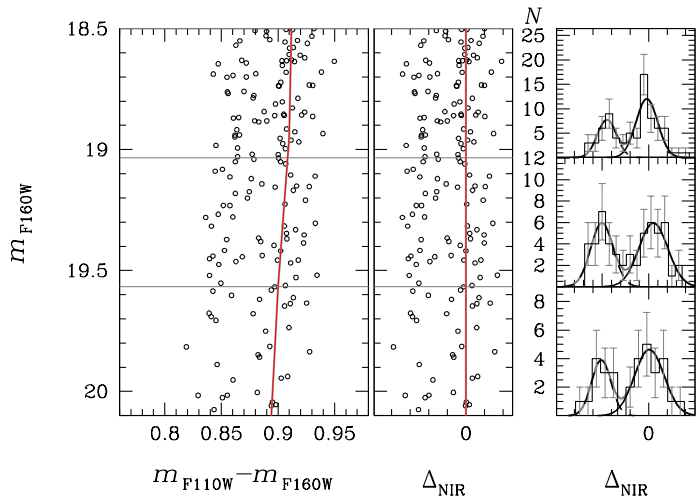}
 \caption{\textit{Left panel}: Zoom of the NIR CMD for M\,4 stars around the bottom of the MS, where the MS split is more evident.  The red line is the ridge line of the $MS_{\rm II}$. Middle panel shows the verticalized MS, while the histograms of the $\Delta$($m_{\rm F110W}-m_{\rm F160W}$) distribution are plotted in right panels for stars in three intervals of F160W magnitude that are separated by the horizontal lines. The gray lines, in each histogram, are the best-fitting bi-Gaussian functions, whose components are represented with continuous and dashed lines.}
 \label{fig:ratio}
\end{figure*}
\end{centering}

Having excluded the possibility that the color spread of VLM MS stars is 
due to photometric errors, we can now
estimate the fraction of stars in each of the two sequences. To do this we applied a procedure that has been widely used by our group (e.g.\, Piotto et al.\, 2007) and is illustrated in Fig.~\ref{fig:ratio} for M\,4. 
We restrict our study to MS stars in the luminosity interval where the split MS is more clearly visible ($18.5<m_{\rm F160W}<20.1$). The left panel of Fig.~\ref{fig:ratio} shows the NIR CMD of these stars while the $m_{\rm F160W}$ versus $\Delta_{\rm NIR}$ diagram is shown in the middle panel. The histogram distribution of $\Delta$($m_{\rm F110W}-m_{\rm F160W}$), which is plotted in the right panel for stars in three F160W intervals 18.50$<m_{\rm F160W}<$19.03, 19.03$<m_{\rm F160W}<$19.57, 19.57$<m_{\rm F160W}<$20.10, is clearly bimodal and has been fitted with the sum of two Gaussians. The fraction of stars in each sequence was estimated by comparing the area under the two Gaussians.  From the upper, the middle, and the fainter histogram, we found that the less-populated MS contains 36\%, 44\%, and 35\% of stars, respectively.  We infer that, on average, 38$\pm$4\% of stars belong to the less-populated $MS_{\rm I}$, while the most-populated $MS_{\rm II}$ is made of the 62$\pm$4\% of stars. The error is estimated as the rms of the three independent estimates divided by the square root of two.

Marino et al.\ (2008), in their spectroscopic study of stellar populations along the RGB of M\,4, concluded that 64$\pm$10\% 
of the RGB stars
are sodium rich and oxygen poor, while the remaining 36$\pm$10\% ones 
have Na and O abundance similar to those of Halo-field stars with the same metallicity. 
We note that the fractions of $MS_{\rm I}$ and $MS_{\rm II}$ stars are identical, at one-sigma level, to the fractions of Na-rich/O-poor and Na-poor/O-rich stars, respectively.
 In the next section, we further investigate the connection between the observed sequences of M-dwarfs and the multiple stellar populations in M\,4.   

\section{Comparison with NGC\,2808}\label{sec:2808}
 The IR channel of the WFC3 camera on {\it HST} revealed multiple sequences of VLM stars in two other massive GCs, namely NGC\,2808 and NGC\,5139 (M12a, Milone\,2012). The case of NGC\,2808 has been studied in 
in particularly large details and therefore it can provide us with
useful information for our study on M\,4. 

The similarity in metallicity between
NGC\,2808 and M\,4 justifies the comparison between the two\footnote{Metallicity estimates of M4 derived from high-resolution spectroscopy include: [Fe/H]=$-$1.18$\pm$0.01 (Ivans et al.\,1999), [Fe/H]=$-$1.07$\pm$0.01 (Marino et al.\,2008), [Fe/H]=$-$1.20$\pm$0.01 (Carretta et al.\,2009), [Fe/H]=$-$1.06$\pm$0.02 (Villanova et al.\,2012), [Fe/H]=$-$1.12$\pm$0.02 (Marino et al.\,2011), [Fe/H]=$-$1.07$\pm$0.01 (Malavolta et al.\, 2013). In the case of NGC\,2808 measurements of [Fe/H] from high-resolution spectra provide: [Fe/H]=$-$1.15$\pm$0.01 (Carretta et al.\,2006), [Fe/H]=$-$1.15$\pm$0.07 (Pasquini et al.\,2011), [Fe/H]=$-$1.22$\pm$0.01 (Marino et al.\,2014).}. 

NGC\,2808 has been widely studied in the context of multiple stellar populations. The upper part of its CMD, from the turn-off down to the MS knee is split in three sequences (Piotto et al.\ 2007), where the middle and the blue MS are highly helium enhanced (up to $Y \sim$0.39 for the bluest MS),   
while the red MS has primordial helium (D'Antona \& Caloi 2004, D'Antona et al.\ 2005, Piotto et al.\, 2007, Milone et al.\,2012b). 
Furthermore, spectroscopic studies have revealed significant star-to-star variations in the light-element abundances (e.g.\, Norris \& Smith 1983, Bragaglia et al.\,2010, Pasquini et al.\,2011, Marino et al.\,2014) with the presence of an extreme Na-O anticorrelation (Carretta et al.\ 2006, 2009).

In the NIR CMD, the three MSs merge together at the luminosity of the MS knee, while at fainter magnitudes, at least two MSs can be identified. A bluer, more populated $MS_{\rm I}$ that contains $\sim65$\% of MS stars, and a $MS_{\rm II}$ with $\sim35$\% of stars. The fractions of stars along $MS_{\rm II}$, and $MS_{\rm I}$ are very similar to the fraction of red-MS stars ($\sim$62\%) and the total fraction of middle-MS and blue-MS stars ($\sim$24+14=38\%, Milone et al.\ 2012b), respectively.

The observed CMD of NGC\,2808 has been compared with appropriate evolutionary models for very low-mass stars and synthetic spectra that account for the chemical composition of the three stellar populations of this clusters (see M12a for details).
It turns out that $MS_{\rm I}$  is associated with the first stellar generation,
which has primordial He, and O-C-rich/N-poor stars, and that  $MS_{\rm II}$,  corresponds to a second-generation stellar population that is enriched in He and N and depleted in C and O.  
The $MS_{\rm I}$ is the faint counterpart of the red MS identified 
by Piotto et al.\,(2007), whereas the $MS_{\rm II}$ corresponds to the lower-mass counterpart of the middle MS and blue MS discussed in the Piotto et al.\,paper.  

\begin{centering}
\begin{figure*}
 \includegraphics[width=13cm]{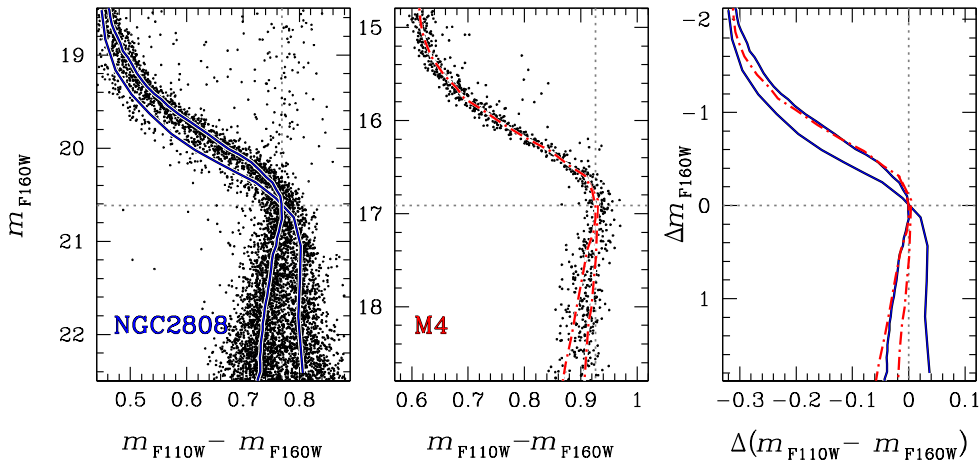}
 \caption{NIR CMD of NGC\,2808 from M12a (left), and of M\,4 (middle).   Blue and red lines superimposed to the CMD of NGC\,2808 and M\,4 respectively, are the fiducial lines of $MS_{\rm I}$ and $MS_{\rm II}$. Dotted lines mark the color and the magnitude of the MS knee. Right panel compares the fiducal lines of NGC\,2808 and M\,4.}
 \label{fig:comparison}
\end{figure*}
\end{centering}

The left panel of Fig.~\ref{fig:comparison} reproduces the $m_{\rm F160W}$ versus $m_{\rm F110W}-m_{\rm F160W}$ CMD of NGC\,2808 from M12a with overlayed the fiducial lines of $MS_{\rm I}$ and $MS_{\rm II}$.
The vertical and horizontal dotted lines mark the color and the magnitude of the reddest point of $MS_{\rm II}$ fiducial. The middle panel shows the same NIR CMD of M\,4 as in Fig.~\ref{fig:cmd} with the fiducials of the two MSs superimposed. 
The fiducial lines of NGC\,2808 and M\,4 MSs are compared in the right panel where we subtracted from the color and the magnitude of each point, the color and the magnitude of the reddest point of the corresponding fiducial ($\Delta$($m_{\rm F110W}-m_{\rm F160W}$) and $\Delta m_{\rm F160W}$). 

The $MS_{\rm I}$ fiducials of M\,4 and NGC\,2808 overlap for almost the entire analyzed $\Delta m_{\rm F160W}$-$\Delta$($m_{\rm F110W}-m_{\rm F160W}$) interval.
 On the contrary, the $MS_{\rm II}$ fiducials exhibit significant differences. 
 Above the MS knee, $MS_{\rm I}$ is significantly redder than $MS_{\rm II}$ in NGC\,2808, while any color difference of the two MSs of M\,4 is small, and $MS_{\rm II}$ is almost superimposed to $MS_{\rm I}$  in our NIR CMD.
Below the MS knee, $MS_{\rm II}$ is redder than $MS_{\rm I}$ in both clusters but the color difference is about 2.5 times larger in the case of NGC\,2808.

\section{Interpreting the split MS}\label{sec:multipop}
\begin{centering}
\begin{figure*}
 \includegraphics[width=13cm]{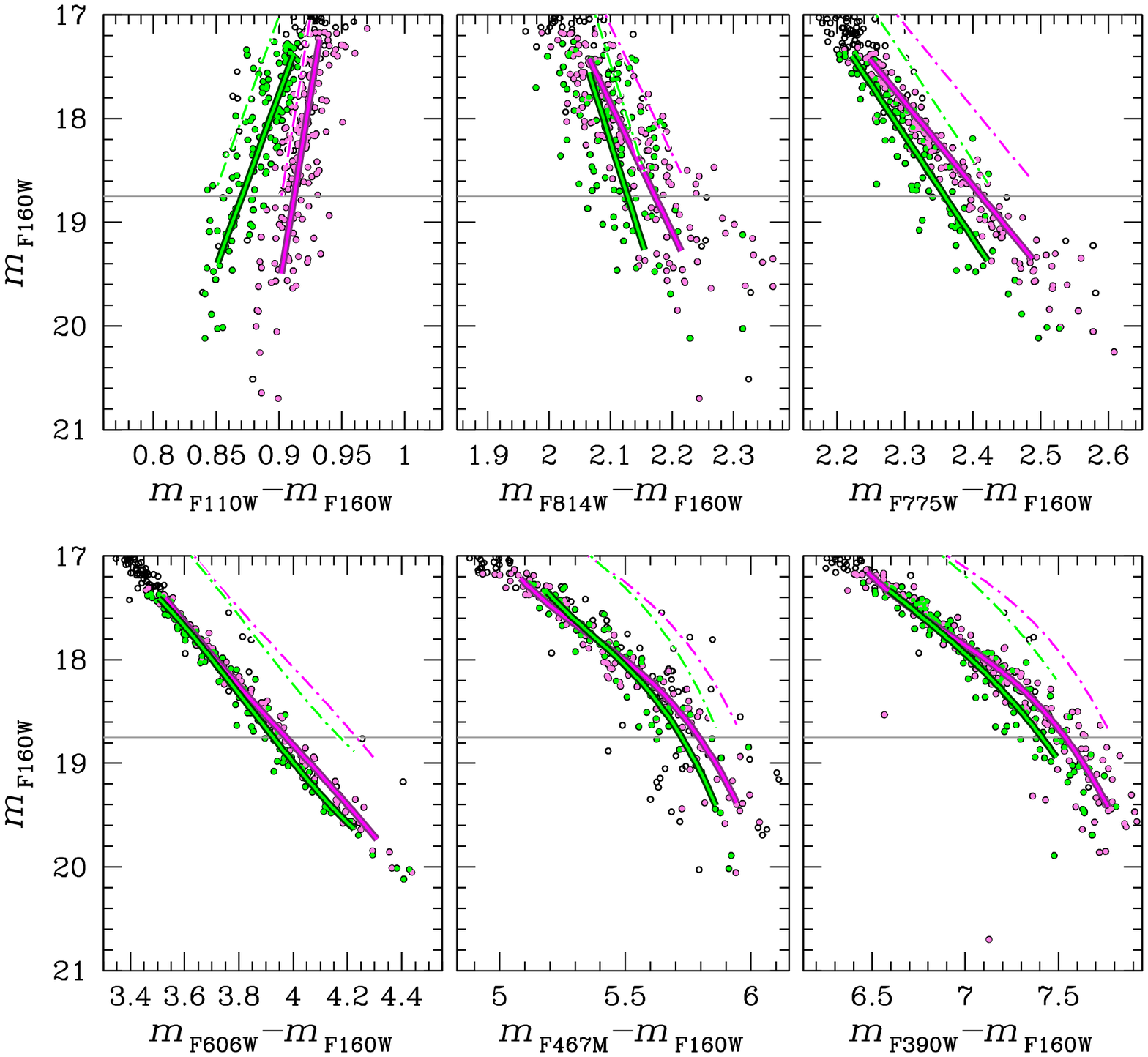}
 \caption{$m_{\rm F160W}$ vs.\, $m_{\rm X}-m_{\rm F160W}$ (X=F110W, F814W, F775W, F606W, F467M, and F390W) CMDs
of M\,4. We colored green and magenta the groups of $MS_{\rm I}$ and $MS_{\rm II}$ stars defined in Fig.~\ref{fig:visual}. Green and magenta continuous lines are the fiducials of 
$MS_{\rm I}$ and $MS_{\rm II}$, respectively, while dashed-dotted lines mark the locus of equal-mass $MS_{\rm I}-MS_{\rm I}$ and $MS_{\rm II}-MS_{\rm II}$ binaries. Grey lines at $m_{\rm F160W}=$18.75 mark the magnitude level where the color separation between $MS_{\rm I}$ and $MS_{\rm II}$ has been calculated (see text for details). }
 \label{fig:multw}
\end{figure*}
\end{centering}

In an attempt to explain the behaviour of $MS_{\rm I}$ and $MS_{\rm II}$ in the CMDs of Fig.~\ref{fig:multw} and to understand the origin of the double MS, we have followed a procedure similar to the one adopted by M12a in their analysis of multiple MSs of VLM stars in NGC\,2808.

 While for NGC\,2808 only NIR photometry was available for VLM stars,
 we can combine visual and IR photometry for M\,4 to get information of the two MSs of M-dwarfs by comparing their relative position in different CMDs. Since we have photometry in seven filters, we can derive six CMDs $m_{\rm F160W}$ vs.\, $m_{\rm X}-m_{\rm F160W}$ (X=F390W, F467M, F606W, F775W, F814W, F110W) that are shown in Fig.~\ref{fig:multw} where we have used for $MS_{\rm I}$ and $MS_{\rm II}$ stars the same color codes introduced in Fig.~\ref{fig:visual}.

In all the cases, $MS_{\rm I}$ is bluer than $MS_{\rm II}$ but the separation changes slightly from one color to another. To quantify the separation of the two sequences we have calculated the color distance between $MS_{\rm I}$ and $MS_{\rm II}$ at $m_{\rm F160W}=$18.75, which is the magnitude corresponding to the horizontal lines in Fig.~\ref{fig:multw}. 
These color differences, $\delta$($m_{\rm X}-m_{\rm F160W}$), are listed in Table~1.

\begin{table*}
\caption{ Color difference 
between $MS_{\rm I}$ and $MS_{\rm II}$ at $m_{\rm F160W}=$18.75.}
\begin{tabular}{cccccc}
\hline\hline
$\delta$($m_{\rm F390W}-m_{\rm F160W}$) & $\delta$($m_{\rm F467M}-m_{\rm F160W}$) &  $\delta$($m_{\rm F606W}-m_{\rm F160W}$) &  $\delta$($m_{\rm F775W}-m_{\rm F160W}$) &   $\delta$($m_{\rm F814W}-m_{\rm F160W}$) &  $\delta$($m_{\rm F110W}-m_{\rm F160W}$) \\
\hline
 $-$0.126$\pm$0.023 & $-$0.069$\pm$0.018 & $-$0.047$\pm$0.010 & $-$0.054$\pm$0.008 & $-$0.054$\pm$0.005 & $-$0.042$\pm$0.008\\
\hline\hline
\end{tabular}\\
\end{table*}

\subsection{The possible effect of binaries}\label{subsec:binaries}
A broadening of the MS could be the result of binaries which are observed in GCs as a spread of the MS. At the distance of M\,4, a binary system will appear as a single point-like source, whose luminosity is equal to the sum of the luminosity of its components.
Equal-mass MS binaries (MS-MS binary) form a sequence that runs almost parallel to the MS, $\sim$0.75 magnitudes brighter.  When the  masses of  the two components  are different, the binary will appear brighter than each component and populate a CMD region between the fiducial line and the equal-mass binary line.
In Fig.~\ref{fig:multw} we have superimposed to each CMD the fiducial lines of $MS_{\rm I}$ and $MS_{\rm II}$ (green and magenta on continuous lines) as well as the fiducial lines of equal-mass $MS_{\rm I}-MS_{\rm I}$ and $MS_{\rm II}-MS_{\rm II}$ binaries (green and magenta dashed-dotted lines). 
The fact that in the $m_{\rm F160W}$ versus $m_{\rm F110W}-m_{\rm F160W}$ CMD all $MS_{\rm II}$ stars lie outside the region between the two green lines demonstrate that $MS_{\rm II}$ cannot be explained with $MS_{\rm I}-MS_{\rm I}$ binaries. Similarly, the evidence that in several CMDs of Fig.~\ref{fig:multw}, (like $m_{\rm F160W}$ versus $m_{\rm F110W}-m_{\rm F160W}$ and $m_{\rm F160W}$ versus $m_{\rm F775W}-m_{\rm F160W}$) most of $MS_{\rm I}$ stars are located outside the region populated by $MS_{\rm II}-MS_{\rm II}$ binaries demonstrates 
that $MS_{\rm I}$ is inconsistent with a population of $MS_{\rm II}-MS_{\rm II}$ binaries.
 
\subsection{The effect of multiple populations}\label{subsec:multipop}
In the previous section, we have demonstrated
that binaries are not responsible of the splitting of the MS.
Here, we investigate the possibility that MS bimodality is due 
to the presence of multiple generations of stars. 
In order to interpret our observations  we need to compare 
 the observed colors of $MS_{\rm I}$ and $MS_{\rm II}$ M-dwarfs
with predictions from theoretical models as done by M12a for NGC\,2808.
Since NGC\,2808 and M\,4 have similar metallicities, for M\,4, we used
the grid of evolutionary models for low- and very low-mass stars adopted by M12a for NGC\,2808. 

These theoretical  models have been calculated by using the same physical inputs described by Cassisi et al.\,(2000) and Pietrinferni et al.\,(2006) and have been converted into the observational domain by integrating the synthetic spectra of the BT-Settl AGSS model atmosphere grid\footnote{http://phoenix.ens-lyon.fr/simulator.} (Allard et al.\,2012) over the IR WFC3 bandpasses. 

 As already discussed in Sect.~\ref{sec:intro}, M\,4 exhibits star-to-star variations in the abundance of several light elements like (C, N, O, Mg, Al, Na), which affect the color of the RGB, the SGB and MS (Marino et al.\,2008, Sbordone et al.\,2011, Milone et al.\,2012c, Cassisi et al.\, 2013). 
Since the literature
model atmospheres used to calculate color-$T_{\rm eff}$ transformation do not account for the light-element abudances
of the two stellar populations of M\,4, we have computed synthetic spectra with the appropriate chemical composition. Specifically we assumed for $MS_{\rm I}$ and $MS_{\rm II}$ the abundances of C, N, O, Mg, Al, and Na as measured for first and second-generation RGB stars by Marino et al.\,(2008, see their Table~6). 

We also assumed that $MS_{\rm II}$ is slightly helium enhanced (Y=0.27) with respect to $MS_{\rm I}$ which has primordial helium content (Y=0.248) as inferred by D'Ercole et al.\,(2010) on the basis of their comparison of chemical-evolution models and measurements of sodium and oxygen in RGB stars.

We have used the ATLAS9 and SYNTHE programs (Kurucz\,2005, Sbordone et al.\,2007) to calculate synthetic spectra of a $MS_{\rm I}$ and $MS_{\rm II}$ star at $m_{\rm F160W}=$18.75.
We assumed $T_{\rm  eff}$=3700K, $\log{g}$=5.3, and microturbulence 1.0, that are the values derived for a M-dwarf with $m_{\rm F160W}=$18.75 from the best-fitting BaSTI isochrones (Pietrinferni et al.\,2004, 2006).
 We accounted for CO, $\rm C_{2}$, CN, OH, MgH, SiH, $\rm H_{2}O$, TiO, VO, and ZrO from Kurucz\,(2005), Partrige \& Schwenke\,(1997), and B.\,Pletz\,(2012, private communication).  Synthetic spectra have been integrated over the transmission curves of the ACS/WFC F606W, F775W, WFC3/UVIS F390W, F467M and F814W and WFC3/IR F110W and F160W filters to derive synthetic colors.

\begin{centering}
\begin{figure*}
\includegraphics[width=8cm]{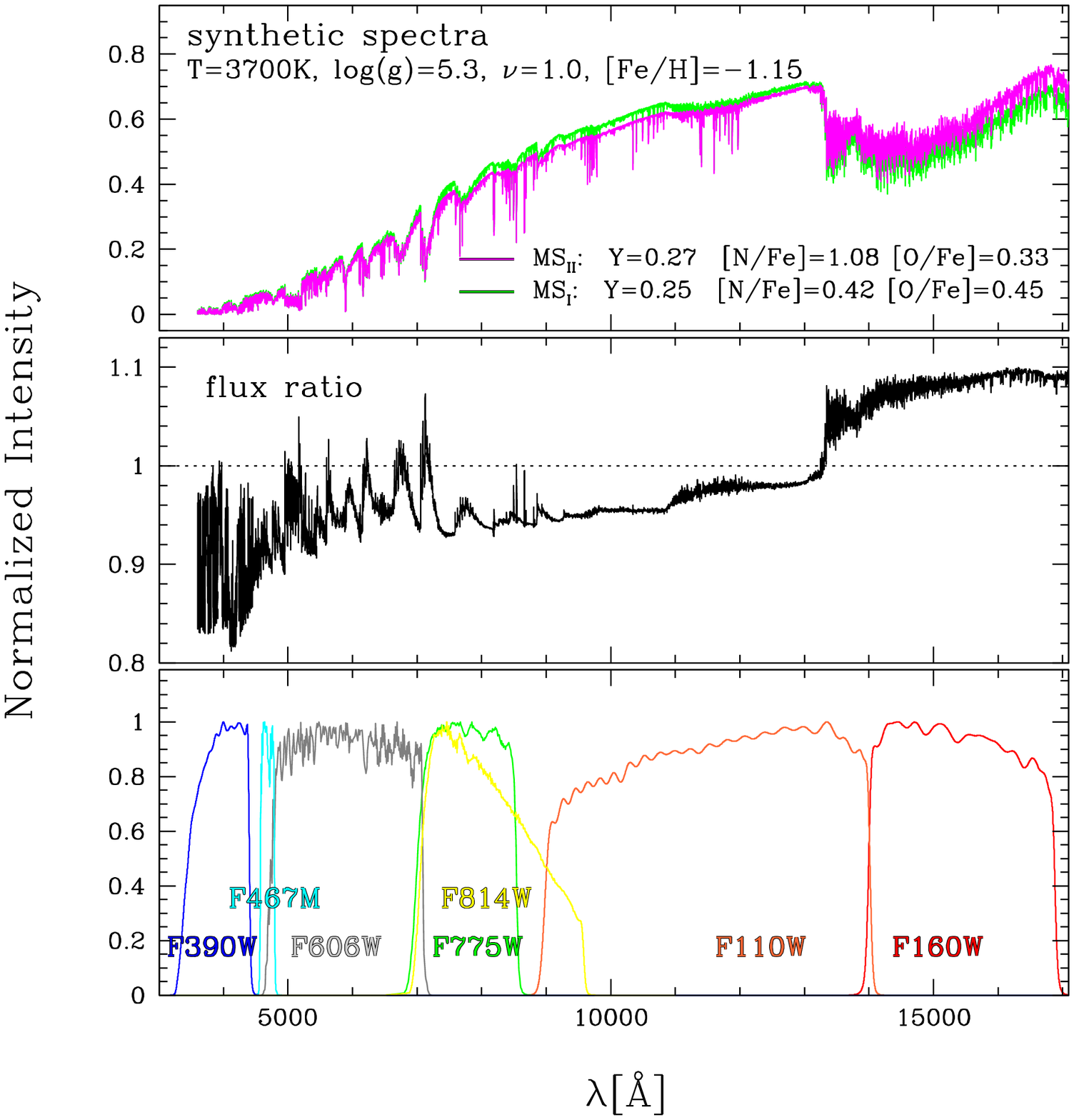} 
\includegraphics[width=8cm]{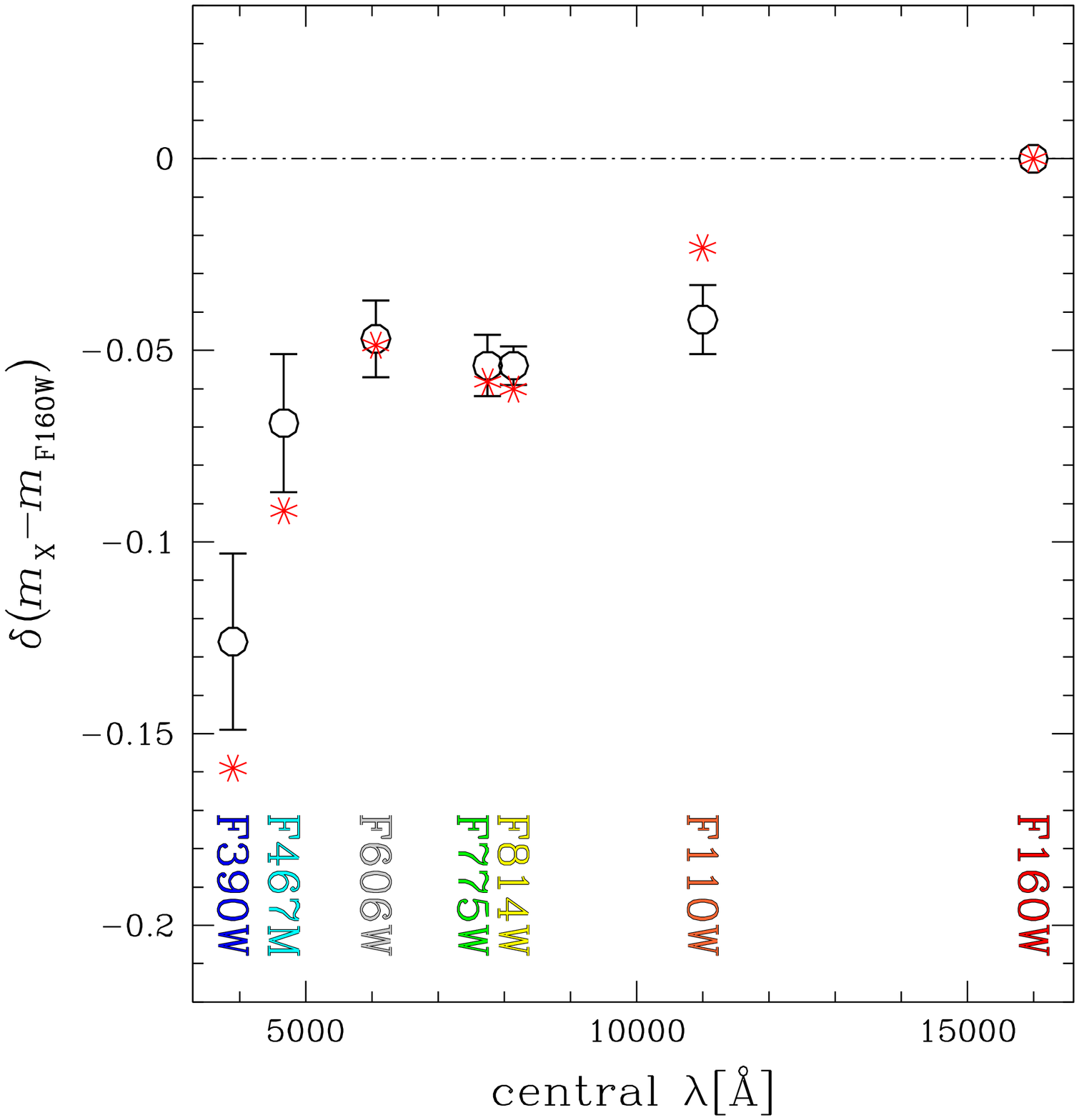}
 \caption{
\textit {Left:} Synthetic spectra for a $MS_{\rm I}$ (green) and $MS_{\rm II}$ (magenta) star at $m_{\rm F160W}=$18.75 (upper panel). Flux ratio (middle panel). Transmission curves of the filters used in this work (lower panel).  
\textit {Right:} $m_{\rm X}-m_{\rm F160W}$ color distance between $MS_{\rm I}$ and $MS_{\rm II}$ at $m_{\rm F160W}=$18.75 (X=F390W, F467M, F606W, F775W, F814W, F110W, and F160W). Observations are represented with black circles, while red asterisks indicate results from synthetic spectra.}
 \label{fig:spettri}
\end{figure*}
\end{centering}

The synthetic spectra of a representative $MS_{\rm I}$ and $MS_{\rm II}$ star $m_{\rm F160W}=$18.75 are compared in the upper-left panel of Fig.~\ref{fig:spettri}, while the flux ratio is plotted on the middle-left. 
The strong absorption at $\lambda>$13000$\AA$, previously observed in the synthetic spectra of $MS_{\rm I}$ star of NGC\,2808, is present in spectra of M\,4 $MS_{\rm I}$ and $MS_{\rm II}$ and is mainly due to ${\rm H_{2}O}$ molecules.
In the bottom-left panel we also show the transmission curve of the ACS/WFC and NIR/WFC3 filters used in the analysis.

In contrast with M\,4 where the difference in [O/Fe] between the first and second stellar population is smaller than 0.2 dex (e.g.\, Marino et al.\,2008), second-generation stars of NGC\,2808 can be depleted in oxygen by up to $\sim$1.3 dex (e.g.\, Carretta et al.\,2006).
The feature at $\lambda>$13000$\AA$ is less evident in the synthetic spectra of $MS_{\rm II}$ stars of NGC\,2808 as a consequence of their extreme O-depletion (see Fig.~4 of M12a).

 The fact that the color separation between $MS_{\rm I}$ and $MS_{\rm II}$ of M\,4 is small when compared to NGC\,2808, seems to be due to the small O-depletion of second-generation stars in M\,4.
As shown in the right panel of Fig.~\ref{fig:spettri}, the trend of the synthetic-color differences is consistent with the observations.  
Figure~\ref{fig:spettri} supports the idea that $MS_{\rm I}$ stars represent a first stellar generation, having primordial helium and high oxygen, while $MS_{\rm II}$ can be associated with a second generation consisting of stars enhanced in He, N, Na and depleted in O. 

\section{Summary}\label{sec:summary}
 NIR infrared photometry obtained with the IR/WFC3 camera of {\it HST} has revealed a double sequence of M-dwarfs in the CMD of the
GC\,M4 and allowed us to extend the study of its stellar populations to the VLM regime.
The NIR CMD of M\,4 shows that, below the knee, the MS splits into two distinct components
 that we designate $MS_{\rm I}$ and $MS_{\rm II}$.
They contain 38$\pm$4\% and  62$\pm$4\% of the stars, respectively.

The comparison of observations and synthetic spectra of M-dwarf which accounts
for the chemical composition of the two stellar populations in M\,4 shows
that the presence  of the $\rm H_{2}O$ molecule on the atmosphere of these cold stars has a strong effect on their $m_{\rm F110W}-m_{\rm F160W}$ color.
The fact that light-element variations are inferred also among fully-convective M-dwarfs provides strong evidence of their primordial origin and demonstrates
that they correspond to different stellar generations.
We associate the
$MS_{\rm I}$ with the first stellar generation, which share the same chemical composition as the Na-poor/O-rich stars observed by Marino et al.\,(2008) along the RGB. $MS_{\rm II}$ corresponds to a second stellar generation and is enhanced in Na, N, and Al and depleted in C, O, and Mg.

We compare the NIR CMDs of M\,4 and NGC\,2808 where multiple sequences of M-dwarfs have been observed by M12a. We find that the color separation between $MS_{\rm I}$  and $MS_{\rm II}$ is significantly larger in NGC\,2808, likely as a consequence of the extreme abundance pattern of its second generations.
 The detection of multiple sequences of VLM stars in M\,4 suggests that this can be a common feature in GCs and not a peculiarity of massive clusters like NGC\,2808 and $\rm \omega$ Centauri. 
NIR observations with WFC3 on board {\it HST} have proven to be a powerful and sensitive tool to 
study the effect of even small abundance differences on the faint, low-mass MS, and therefore an additional observational windows for the study of
multiple stellar populations in GCs down to the lowest-mass MS stars.

 \section*{acknowledgments}
\small
APM and HJ acknowledge the financial support from the Australian Research 
Council through Discovery Project grant DP120100475.
SC is grateful for financial support  from PRIN-INAF 2011 "Multiple Populations in Globular Clusters: their  role in the Galaxy assembly" (PI: E.\,Carretta), and from PRIN MIUR 2010-2011,  project \lq{The Chemical and Dynamical Evolution of the Milky Way and Local Group Galaxies}\rq, prot. 2010LY5N2T (PI: F.\,Matteucci). LRB and GP acknowledge PRIN-INAF 2012 funding under the project entitled: \lq{The M4 Core Project with Hubble Space Telescope}.

\bibliographystyle{aa}

\end{document}